\documentclass[aps,prl,twocolumn,superscriptaddress,floatfix]{revtex4}

\usepackage{graphicx}
\usepackage{dcolumn}
\usepackage{bm}
\usepackage{amsmath}
\usepackage{amssymb}
\usepackage{color}

\begin{document}

\renewcommand{\vec}[1]{\mbox{\boldmath $#1$}}


\title{Emergence of spin-orbit order in the spinel CuCr$_2$O$_4$}


\author{K. Tomiyasu}
\email[Electronic address: ]{tomiyasu@tohoku.ac.jp}
\affiliation{Department of Physics, Tohoku University, Aoba, Sendai 980-8578, Japan}
\author{S. Lee}
\affiliation{Institute of Materials Structure Science, High Energy Accelerator Research Organization (KEK), Tokai, Ibaraki 391-1106, Japan}
\author{H. Ishibashi}
\affiliation{Department of Physical Science, Osaka Prefecture University, Sakai, Osaka 599-8531, Japan}
\author{Y. Takahashi}
\affiliation{Department of Physics, Tohoku University, Aoba, Sendai 980-8578, Japan}
\author{T. Kawamata}
\affiliation{Department of Applied Physics, Tohoku University, Sendai 980-8579, Japan}
\author{Y. Koike}
\affiliation{Department of Applied Physics, Tohoku University, Sendai 980-8579, Japan}
\author{T. Nojima}
\affiliation{Institute for Materials Research, Tohoku University, Sendai 980-8577, Japan}
\author{S. Torii}
\affiliation{Institute of Materials Structure Science, High Energy Accelerator Research Organization (KEK), Tokai, Ibaraki 391-1106, Japan}
\author{T. Kamiyama}
\affiliation{Institute of Materials Structure Science, High Energy Accelerator Research Organization (KEK), Tokai, Ibaraki 391-1106, Japan}
\affiliation{Department of Materials Structure Science, SOKENDAI, Tokai, Ibaraki, 319-1106, Japan}


\date{\today}

\begin{abstract}
We determined the magnetic structure of CuCr$_2$O$_4$ using neutron diffraction and irreducible representation analysis. 
The measurements identified a new phase between 155 K and 125 K as nearly collinear magnetic ordering in the Cr pyrochlore lattice. Below 125 K, a Cu-Cr ferrimagnetic component develops the noncollinear order. 
Along with the simultaneously obtained O positions and the quantum effect of spin-orbit coupling, the magnetic structure is understood to involve spin-orbit ordering, accompanied by an appreciably deformed orbital of presumably spin-only Cu and Cr. 
\end{abstract}


\maketitle

%
%
The concept of frustration has provided fertile sources for various exotic ground states in matter since Pauling proposed water ice~\cite{Pauling_1935}. In frustrated magnets, not all classical-spin pairs can be arranged antiferromagnetically on a triangular or tetrahedral lattice, which gives rise to an inherent macroscopic degeneracy~\cite{Wannier_1950, Anderson_1956, Anderson_1987}. Therefore, novel orders often emerge via spin-orbital-lattice coupling, such as the underlying structures for multi-ferroics, orbital ordering, and spin-orbit molecule organization~\cite{Kimura_2003, Book_Ramesh_2012, Lee_2010, Senn_2012}. 

Spinels are typical spin-frustrated materials. For example, in $A$Cr$_2$O$_4$, the $A$ = Co and Mn systems exhibit both ferromagnetism and ferroelectrics on the basis of their conical spin structure~\cite{Yamasaki_2006, Menyuk_1964, Hastings_1962, Tomiyasu_2004}. The $A$ = Mg, Zn, Cd, Hg systems exhibit a magnetization plateau under ultrahigh magnetic field and molecular spin excitations, both of which are caused by spin-lattice coupling and are tightly related to topological physics~\cite{Ueda_2005, Miyata_2011a, Miyata_2011b, Tomiyasu_2008, Tomiyasu_2013, Watanabe_2012, Shannon_2010, Tchernyshyov_2002, Conlon_2010, Mizoguchi_2017, Paddison_2015}. 

\begin{figure}[htbp]
\begin{center}
\includegraphics[width=0.8\linewidth, keepaspectratio]{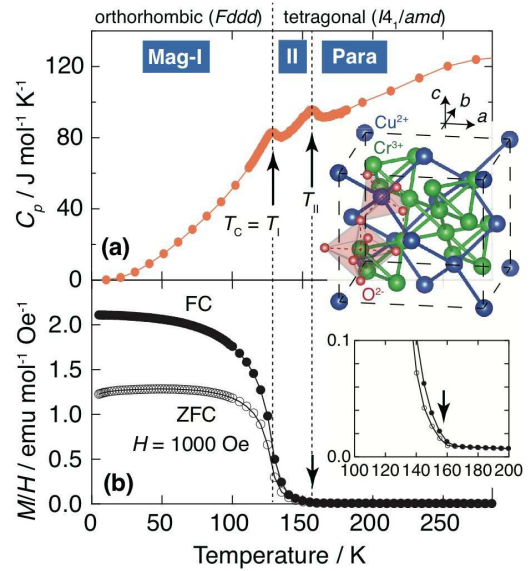}
\end{center}
\caption{\label{fig:intro} (Color online)
Temperature dependence of (a) heat capacity $C_{p}$ and (b) magnetic susceptibility $\chi=M/H$. The upper inset illustrates the crystal structure. 
The lower inset magnifies the low magnetic susceptibility range at approximately 155 K.}
\end{figure}

The $A$ = Cu system, CuCr$_2$O$_4$, also belongs to this series. The crystal structure is depicted in the inset of Fig.~\ref{fig:intro}. The magnetic Cr$^{3+}$ ($d^3$), which is octahedrally surrounded by O$^{2-}$, is described by an isotropic spin $S=3/2$ without an orbital degree of freedom, and constructs the pyrochlore sublattice as the source of strong frustration. The magnetic Cu$^{2+}$ ($d^9$) resides in a tetrahedral ligand field. Its Jahn-Teller activity results in a large tetragonal lattice contraction by $1-c/a \simeq 0.1$ below $T_{\rm JT} \simeq 850$ K~\cite{Ye_1994}, in which Cu$^{2+}$ forms $S=1/2$ without an orbital degree of freedom. As the temperature further decreases, CuCr$_2$O$_4$ undergoes simultaneous transitions of the ferrimagnetic order and slight tetragonal-to-orthorhombic lattice distortion of $1-a/b \simeq 5 \times 10^{-4}$ (space group $I4_{1}/amd$ to $Fddd$) at $T_{\rm C} \simeq 125$ K~\cite{Suchomel_2012}. The neutron-diffraction research began in 1957 and proposed the so-called Yafet-Kittel triangular type of magnetic structure~\cite{Prince_1957}, which was refined by a modern neutron diffractometer and Rietveld analysis~\cite{Reehuis_2015}. 

However, a new anomaly was recently reported at approximately 155 K in addition to the ferrimagnetic transition at $T_{\rm C}$ in specific heat and magnetization measurements~\cite{Suchomel_2012}, although high-resolution synchrotron X-ray diffraction detects no disruption of the symmetry in the crystal structure near this temperature~\cite{Suchomel_2012}. The use of neutron diffraction to investigate the magnetic structure in the temperature range between 155 K and $T_{\rm C}\simeq125$ K has not yet been reported. Thus, the origin of this range remains a mystery. 
Furthermore, the Yafet-Kittel magnetic structure at the lowest temperature is yet to be determined, as the extremely small orthorhombic $ab$ distortion has been unresolved and the orthorhombic magnetic symmetry has not been examined by neutron diffraction. 
This prompted us to determine the magnetic structures in the two phases by combining state-of-the-art high-resolution time-of-flight neutron diffractometry and irreducible representation analysis~\cite{Book_Izyumov_1991, Book_Kovalev_1993}. Further, by using information of the positions of the oxygen atoms, which were simultaneously obtained and at which neutron diffraction also excels, we clarify the origin of magnetic ordering. 

%
%
$Experiments.$-- 
Neutron diffraction experiments were performed on the Super-High-Resolution Powder Diffractometer (SuperHRPD) at the MLF of the J-PARC spallation neutron source in Japan~\cite{Torii_2011, Torii_2014}. Data were recorded using the 90$^{\circ}$-middle and 172$^{\circ}$-backward banks of position-sensitive detectors, at resolutions of $\Delta Q/Q \sim 4\times10^{-3}$ and $4\times10^{-4}$, respectively. Approximately 3 g of the sample was sealed in a 6-mm-diameter thin V cylinder with He exchange gas, which was positioned under the cold head in a He closed-cycle refrigerator. The crystal and magnetic structures were analyzed using the FullProf and SARA$h$ software~\cite{Rodriguez_Carvajal_1993, Wills_2000}.
A powder sample of CuCr$_2$O$_4$ was synthesized by a solid-state reaction method, in which a stoichiometric mixture of CuO and Cr$_2$O$_3$ was ground, followed by the calcination at 1000 $^{\circ}$C for 24 h and 1200 $^{\circ}$C for 24 h in an O$_2$-gas flow with intermediate grinding and pelletizing. 
The heat capacity was measured using the Physical Properties Measurement System (PPMS) at the Department of Applied Physics, Tohoku University, Japan. Direct-current magnetization was measured using superconducting quantum interference device (SQUID) magnetometers at the Center for Low-Temperature Science at this university. 


%
%
$Results.$-- 
First, before starting with the neutron experiments, we measured the temperature dependence of the heat capacity. As shown in Fig.~\ref{fig:intro}(a), two phase transitions at approximately 155 K and 125 K are clearly observed. As the temperature decreases, the magnetic susceptibility begins to rapidly increase below 155 K and shows a maximum slope at approximately 125 K, as shown in Fig.~\ref{fig:intro}(b). This demonstrates that CuCr$_2$O$_4$ has two magnetic transitions, which are also confirmed by the magnetization curve measurements~\cite{SM}. Hereinafter, we refer to the three temperature ranges as the Para, Mag-II, and Mag-I phases and the two phase-transition temperatures as $T_{\rm II}\simeq155$ K and $T_{\rm I}=T_{\rm C}\simeq125$ K. 

\begin{figure}[htbp]
\begin{center}
\includegraphics[width=0.85\linewidth, keepaspectratio]{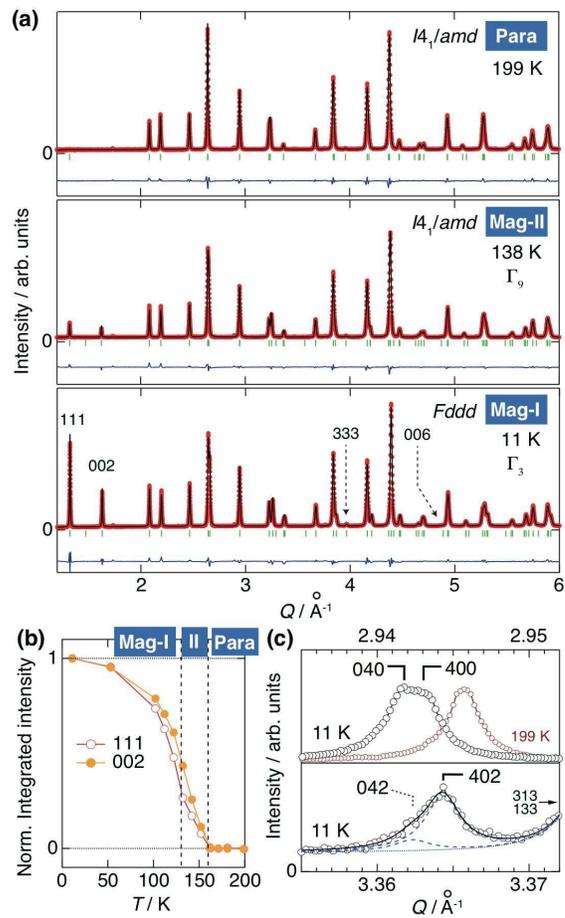}
\end{center}
\caption{\label{fig:neutron} (Color online)
(a) Powder neutron diffraction patterns measured in the middle bank in the three phases. 
The fitting curves were obtained for the crystal and magnetic structures described by the displayed symmetries. The reliability factor $R_{\rm wp}$ was 6.3, 4.1, and 6.7{\%}, respectively. (b) Temperature dependence of the 111 and 002 integrated intensities normalized at 11 K. 
(c) Selected reflections recorded at 11 K in the high-resolution backward bank. The curves in the lower panel are the fitting results obtained in the $\Gamma_{3}$ magnetic structure (see the text). }
\end{figure}

Next, we performed the neutron diffraction experiments. The crystallographic $a$, $b$, and $c$ axes and Miller indices are represented in the pseudo-cubic/orthorhombic $F$-lattice notation in the entire paper, as shown in the upper inset of Fig.~\ref{fig:intro}. Figure~\ref{fig:neutron}(a) shows the patterns recorded in the middle bank in these three phases. As the temperature decreases from the Para phase, the low-$Q$ 111 and 002 reflections intensify in the Mag-II and Mag-I phases. In contrast, their high-$Q$ equivalent 333 and 006 reflections are very weak even in the lowest-temperature Mag-I phase, indicating that the 111 and 002 reflections are magnetic in origin. These magnetic reflections are indexed with the propagation vector $\vec{k} = (0,0,0)$. 

Figure~\ref{fig:neutron}(b) shows the temperature dependence of the normalized integrated intensity, which represents an increase in the Mag-II phase in concurrence with the aforementioned heat capacity and magnetic susceptibility data. 

Figure~\ref{fig:neutron}(c) shows the typical reflections recorded in the high-resolution backward bank. As shown in the upper panel, the 040 and 400 reflections can be distinguished in comparison with the Para-phase profile, confirming that the resolution was sufficiently high to discuss the effects of the slight $ab$ orthorhombic distortion of $ca.$ $5 \times 10^{-4}$. Furthermore, the lower panel shows the 042 and 402 reflections, which are extinct under the nuclear/crystallographic reflection conditions and therefore are purely magnetic. The magnetic intensity of the 042 reflection is much weaker than that of 402. 

%
%
$Analysis.$-- 
Now we analyze the magnetic structure. In the Mag-II phase, the possible $\vec{k}=(0,0,0)$ magnetic structures in the tetragonal $I4_{1}/amd$ space group are described by the irreducible representations, $\Gamma_{\rm Cu} = \Gamma_{3} + \Gamma_{6} + \Gamma_{9}^{2} + \Gamma_{10}^{2}$ at the Cu position and $\Gamma_{\rm Cr} = \Gamma_{1} + 2\Gamma_{3} + \Gamma_{5} + 2\Gamma_{7} + 3\Gamma_{9}^{2}$ at the Cr position, where the coefficients denote the number of basis vectors and the superscript 2 the two-dimensional set of basis vectors~\cite{Wills_2000}. Among them, the $\Gamma_{3}$ and $\Gamma_{9}$ representations are common to both $\Gamma_{\rm Cu}$ and $\Gamma_{\rm Cr}$. Furthermore, $\Gamma_{3}$ is precluded as the 002 magnetic reflection is forbidden contrary to the experimental data. Hence, $\Gamma_{9}$ is the most probable. 

\begin{figure}[htbp]
\begin{center}
\includegraphics[width=0.95\linewidth, keepaspectratio]{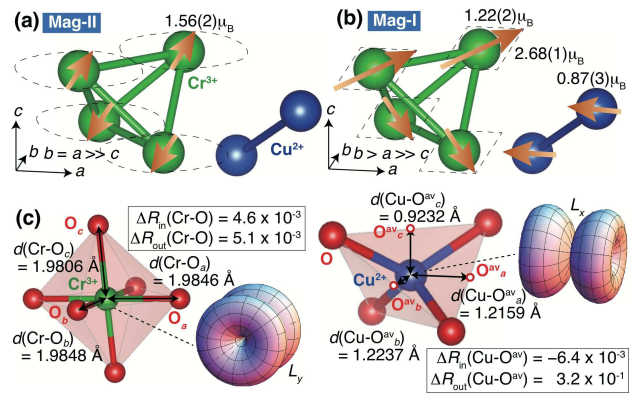}
\end{center}
\caption{\label{fig:magstr} (Color online)
Magnetic unit cells in the Mag-II (a) and Mag-I (b) phases. All the magnetic moments lie in the $ab$ plane. 
(c) CrO$_6$ and CuO$_4$ polyhedra and the expected orbital shapes of unquenched $L$ in Mag-I. The parameters of the in-plane and out-of-plane ratios are defined by 
%
$\Delta R_{\rm in}({\rm Cr\mathchar`-O}) = 1 - d({\rm Cr\mathchar`-O}_{a})/d({\rm Cr\mathchar`-O}_{b})$, 
$\Delta R_{\rm out}({\rm Cr\mathchar`-O}) = 1 - d({\rm Cr\mathchar`-O}_{c}) / 0.5\{d({\rm Cr\mathchar`-O}_{a}) + d({\rm Cr\mathchar`-O}_{b}) \}$; 
$\Delta R_{\rm in}({\rm Cu\mathchar`-O^{\rm av}}) = 1 - d({\rm Cu\mathchar`-O}^{\rm av}_{a})/d({\rm Cu\mathchar`-O}^{\rm av}_{b})$, 
$\Delta R_{\rm out}({\rm Cu\mathchar`-O^{\rm av}}) = 1 - d({\rm Cu\mathchar`-O}^{\rm av}_{c}) / 0.5\{d({\rm Cu\mathchar`-O}^{\rm av}_{a}) + d({\rm Cu\mathchar`-O}^{\rm av}_{b}) \}$, 
where O$^{\rm av}$ denotes the virtual midpoint between two oxygen atoms in CuO$_4$, as shown by the red open circles. }
\end{figure}

The result of Rietveld fitting performed for the $\Gamma_{9}$ magnetic structure, which is shown in the middle panel in Fig.~\ref{fig:neutron}(a), is satisfactorily consistent with the experimental data, and the obtained magnetic structure is shown in Fig.~\ref{fig:magstr}(a). Although the Cu ordered moment is not perceptible ($ \lesssim 0.1 \mu_{\rm B}$), the Cr ordered moment is 1.56$\mu_{\rm B}$. The spatial correlation is antiferromagnetic along the $c$-axis and ferromagnetic in the $ab$ plane. The moment direction is collinear and lies in the $ab$ plane, where $a=b$. 

However, the in-plane direction cannot be determined in powder diffraction and also cannot be symmetrically restricted owing to the two-dimensionality ($XY$ type) of $\Gamma_{9}$. Thus, in Fig.~\ref{fig:magstr}(a), we select the $a=b$ axis as the easy axis so as to continuously connect to the following Mag-I magnetic structure [Fig.~\ref{fig:magstr}(b)], as explained below. 

%

The magnetic structure in the Mag-I phase was analyzed similarly. 
In the $Fddd$ space group, the Cu and Cr common irreducible representations are $\Gamma_{3}$, $\Gamma_{5}$, and $\Gamma_{7}$. Among them, $\Gamma_{7}$ is ruled out, as the 002 magnetic reflection is forbidden, contrary to our experimental data. Thus, either the $\Gamma_{3}$ or $\Gamma_{5}$ representation is expected to describe the overall magnetic structure of CuCr$_2$O$_4$. 

The lower panel in Fig.~\ref{fig:neutron}(a) shows the result of the Rietveld fitting performed for the $\Gamma_{3}$ magnetic structure, which corresponds to the experimental data. Figure~\ref{fig:magstr}(b) shows the obtained $\Gamma_{3}$ magnetic structure. This noncollinear structure consists of the collinear ferrimagnetic (F) component of the anti-parallel Cu and Cr moments along the slightly shorter $a$-axis in addition to the Mag-II collinear antiferromagnetic (AF) component of Cr along the slightly longer $b$-axis. This is consistent with the rapid increase in the F-component and the slight $ab$ lattice distortion simultaneously occurring below $T_{\rm I}$~\cite{Suchomel_2012}, suggesting that the ordering of the F-component needs to distinguish between $a$ and $b$ to select either of them as its easy axis. Furthermore, the AF-component belongs to the aforementioned $XY$ magnetic structure in the Mag-II phase, which enables us to make the natural connection between them. 

However, the $\Gamma_{5}$ magnetic structure, in which the slightly inequivalent $a$- and $b$-axes are only switched, gives a virtually identical fitting curve for the middle-bank data. Thus, to distinguish between the $a$- and $b$-directions, we use the high-resolution bank data of the 042 and 402 magnetic reflections [lower panel of Fig.~\ref{fig:neutron}(c)]. These reflections arise from only the AF-component, because the F-component is extinct as is the case with the crystallographic reflection conditions. Furthermore, the neutron magnetic reflection intensity is proportional to the factor of $\sin^{2}\alpha$, where $\alpha$ denotes the angle between the $hkl$ scattering vector and collinear AF magnetic moment. Hence, the intensity ratio of 042 and 402 is expected to be 1:5 and 5:1 for the $\Gamma_{3}$ and $\Gamma_{5}$ structures, respectively, and the acceptable fitting was obtained for the $\Gamma_{3}$ structure, as shown in the lower panel of Fig.~\ref{fig:neutron}(c). Thus, $\Gamma_{3}$ [Fig.~\ref{fig:magstr}(b)] is selected as the final solution. 

In this way, we determined the magnetic structures in the Mag-II and Mag-I phases. Further details of the analysis and refined structural parameters are summarized in the Supplementary Material ~\cite{SM}, where further details of our study of the positions of the ligand oxygen atoms also appear. Figure~\ref{fig:magstr}(c) shows the CrO$_6$ and CuO$_4$ polyhedra together with the characteristic lengths and ratios, obtained in the magnetic structure found at the lowest temperature of 11 K. Surprisingly, the shapes of the polyhedra are quite different from the unit cell in terms of their in-plane orthorhombicity and out-of-plane tetragonality; despite $\Delta R_{\rm in}({\rm cell}) = 5.4 \times 10^{-4} \ll \Delta R_{\rm out}({\rm cell}) = 1.3 \times 10^{-1}$ (Jahn-Teller), 
(1) $\Delta R_{\rm in}({\rm Cr\mathchar`-O}) = 5.1 \times 10^{-3}$ is close to $\Delta R_{\rm out}({\rm Cr\mathchar`-O}) = 4.6 \times 10^{-3}$ and much larger than $\Delta R_{\rm in}({\rm cell})$, 
(2) $\Delta R_{\rm in}({\rm Cu\mathchar`-O^{\rm av}}) = -6.4 \times 10^{-3}$ is opposite in sign and fairly large compared to $\Delta R_{\rm in}({\rm cell})$, where all the $\Delta R$ and O$^{\rm av}$ are defined in Fig.~\ref{fig:magstr}(c). These facts indicate that the orbital clouds of the presumably spin-only Cr and Cu are appreciably deformed and expanded along the in-plane $b$- and $a$-axes, respectively. Furthermore, these axes are identical to the moment directions of the Cr AF-component and Cu-Cr F-component, respectively. Thus, the magnetic structure of CuCr$_2$O$_4$ is understood to be spin-orbit-lattice mixed order. 

%
%
$Discussion.$--
The origin of the spin-orbit-lattice order is understood as follows. Shannon {\it et al.} calculated for a pyrochlore lattice that the antiferromagnetic first-neighbor interaction $J_1$ causes the total spin to be zero in a first-neighbor tetrahedron, the spin-lattice coupling $E_{\rm sl}$ generates the spin collinearity, and $J_1$ and $E_{\rm sl}$ induce the robust local unit of the collinear first-neighbor tetrahedron ($uudd$-type, where $u/d$ denotes up/down spin)~\cite{Shannon_2010}, which is the same as the Mag-II unit cell [Fig.~\ref{fig:magstr}(a)]. In fact, substantial $J_{1}$ and $E_{\rm sl}$ were recognized in magnetization plateau studies under an ultrahigh magnetic field for isomorphic ZnCr$_{2}$O$_{4}$~\cite{Miyata_2011a, Miyata_2011b}. 
In addition, CuCr$_2$O$_4$ exhibits a large tetragonal lattice contraction of 10{\%}, which is absent in ZnCr$_{2}$O$_{4}$ and the theory, suggesting that $|J_{1c}| > |J_{1ab}|$. Therefore, the Cr spins are considered to prefer an antiferromagnetic arrangement along the contracted $c$-axis, which also coincides with the Mag-II structure. Below $T_{\rm I}$, the Neel interaction ($J_{\rm Cu-Cr}$) generates the canted Mag-I structure [Fig.~\ref{fig:magstr}(b)]. 

The remaining issue is the orbital characteristics coupled to the spin anisotropy. As a mechanism for spin-only anisotropy, the quantum effect of second-order spin-orbit coupling (SOC) between $d\epsilon$ and $d\gamma$ was proposed for Fe$^{2+}$ in Fe$_{1-x}$Mn$_{x}$Cr$_2$O$_4$ and Fe$_{1+x}$Cr$_{2-x}$O$_4$~\cite{Ohtani_2010, Ma_2014}. Here, extending this idea, we try to explain both the spin and orbital electronic states in CuCr$_2$O$_4$. 

\begin{figure}[htbp]
\begin{center}
\includegraphics[width=0.8\linewidth, keepaspectratio]{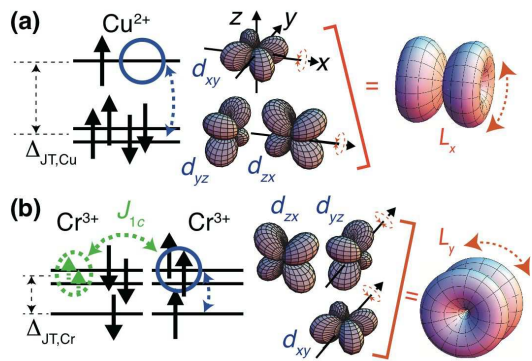}
\end{center}
\caption{\label{fig:discussion} (Color online)
Schematic representation of spin-orbit states in Cu$^{2+}$ (a) and Cr$^{3+}$ (b). Only the $d\epsilon$ orbital is shown and the $d\gamma$ orbital is omitted; the proper orbital occupations are described by energetically low $(d\gamma)^{4}$ and high $(d\epsilon)^{5}$ for Cu$^{2+}$ and low $(d\epsilon)^{3}$ and high $(d\gamma)^{0}$ for Cr$^{3+}$. The tetragonal Jahn-Teller splitting energy $\Delta_{\rm JT}$ is much smaller than the crystal-field energy between $d\epsilon$ and $d\gamma$, which enhances the second-order SOC effect (blue vertical arrows) and generates the $L$ characteristics (orange circular arrows). }
\end{figure}

Figure~\ref{fig:discussion}(a) depicts the $d$-electron state of Cu$^{2+}$. The triply degenerate $d\epsilon$ orbital with a hole is known to recover the orbital angular momentum of $L=1$ by the first-order relativistic SOC, as in Co$^{2+}$ and Ir$^{4+}$~\cite{Kanamori_1957, Kim_2008, Kim_2009}. Although this is not the case with Cu$^{2+}$, the $d_{zx}$ or $d_{yz}$ electron can transit to an empty $d_{xy}$ orbital and return via second-order SOC. This perturbation process partially modulates the $|d_{zx}\rangle$ state to the linear combination $|d_{zx}\rangle + \mathrm{i} \delta |d_{xy}\rangle$ state, in which the spatial rotation of the two orbitals around the $x$-axis generates orbital angular momentum $L_x$. Thus, the Cu$^{2+}$ spin prefers the $x$-direction, along which CuO$_4$ is expanded in accordance with the shape of the $L_x$ orbital, which coincides with the experimental structures in the Mag-I phase [Fig.~\ref{fig:magstr}]. 

Figure~\ref{fig:discussion}(b) depicts the state of Cr$^{3+}$. In this case, it seems to be purely spin-only without a hole in the $d\epsilon$ orbital. This would be true in the Para phase. Upon magnetic ordering, however, the AF direct $J_{1c}$ process leaves behind a slight hole in the $d_{yz}$ and $d_{zx}$ orbitals. Using this hole, Cr$^{3+}$ could activate the second-order SOC mechanism and attain $L_{x}$ and $L_{y}$ characteristics by quantum linear combination. Experimentally, too, the Cr moment lies in the $ab$-plane in both the Mag-II and Mag-I phases. By selecting $L_y$, both the AF moment and CrO$_6$ expansion along the $b$-axis, observed in the Mag-I phase, also arise [Fig.~\ref{fig:magstr}]. 

%

Thus, we arrive at the picture in which the spin-orbit-lattice order is underlain by the Cu ferro-orbital order of $L_x$, Cr ferro-orbital order of $L_y$, and the overall ferri-like-orbital order. Furthermore, we note that the second-order SOC effect occurs inside the energetically proximate $d\epsilon$ for Cu$^{2+}$ and Cr$^{3+}$. Therefore, the expressed orbital characteristics are expected to have been appreciably enhanced. 

%
%
$Summary.$---
We determined the magnetic structures in the Mag-II and Mag-I phases in CuCr$_2$O$_4$ by using a combination of time-of-flight high-resolution neutron diffraction and irreducible representation analysis. 
In the context of the O positions and the quantum effect of SOC, these magnetic structures are understood to be of spin-orbit-lattice mixed ordering, which is expressed in presumably spin-only systems. Our results enable advanced spin-orbit-lattice physics to be expanded in new directions in future. 

%
%
%
%
\acknowledgments
We thank Mr. M. Shioya for assisting with the reduction of neutron data and Dr. K. Aoyama for the discussion about the theory. The neutron experiments were performed with the approval of J-PARC (2014B0261). This study was financially supported by MEXT and JSPS KAKENHI (JP17H06137, JP15H03692) and by the FRIS Creative Interdisciplinary Research Program at Tohoku University. 
K.T. and S.L. equally contributed to this work.

\appendix
\begin{widetext}
\section{Supplementary Material}
\subsection{I. Details of magnetic structure analysis}
Irreducible representation (IR) analysis can classify the possible magnetic structures of Cu and Cr sites and a magnetic structure is normally described by a single IR~\cite{Book_Izyumov_1991, Book_Kovalev_1993}. In the case of the tetragonal Mag-II phase, the space group $I4_{1}/amd$ and magnetic propagation vector $\vec{k}=(0, 0, 0)$ give $\Gamma_{\rm Cu} = \Gamma_{3} + \Gamma_{6} + \Gamma_{9}^{2} + \Gamma_{10}^{2}$ and $\Gamma_{\rm Cr} = \Gamma_{1} + 2\Gamma_{3} + \Gamma_{5} + 2\Gamma_{7} + 3\Gamma_{9}^{2}$. Hence, it is sufficient to investigate only $\Gamma_{\rm Cu-Cr} = \Gamma_{\rm Cu} \cap \Gamma_{\rm Cr} = \{ \Gamma_{3}, \Gamma_{9} \}$. However, the Cu ordered moment is experimentally imperceptible, as reported in the main text. Thus, to be safe, we widely examined all the IRs included in $\Gamma_{\rm Cr}$. 
Table~\ref{tab:BVs_138K} summarizes the basis vectors. 
Among the IRs, as shown in Fig.~\ref{fig:SI_neutron_138K}, only the $\Gamma_{9}$ magnetic structure generates the 002 reflection and is in agreement with the experimental data, leading to the magnetic structure shown in the main text. 

On the other hand, in the orthorhombic Mag-I phase, both the Cu and Cr ordered moments have substantial magnitudes (almost full values of 1$\mu_{\rm B}$ and 3$\mu_{\rm B}$, respectively). This verified the need to focus on $\Gamma_{\rm Cu-Cr}$ and we finally selected the $\Gamma_{3}$ magnetic structure, as explained in the main text. Table~\ref{tab:BVs_011K} summarizes the common basis vectors; the space group $Fddd$ and $\vec{k}=(0, 0, 0)$ give  
$\Gamma_{\rm Cu} = \Gamma_{3} + \Gamma_{4} + \Gamma_{5} + \Gamma_{6} + \Gamma_{7} + \Gamma_{8}$,  
$\Gamma_{\rm Cr} = 3\Gamma_{1} + 3\Gamma_{3} + 3\Gamma_{5} + 3\Gamma_{7}$, 
and hence, $\Gamma_{\rm Cu-Cr} = \{ \Gamma_{3}, \Gamma_{5}, \Gamma_{7} \}$.

\begin{table*}[htbp]
\caption{\label{tab:BVs_138K} Basis vectors of magnetic structures represented by $\Gamma_{\rm Cr}$ used for the Mag-II phase. The space group is described by $I4_{1}/amd$ (No.~141, origin 2) and the magnetic propagation vector $\vec{k}=(0, 0, 0)$. For convenience, the $I$-lattice notation is used in this table. The parameters of $S_{A,x}$, $S_{A,y}$, $S_{A,z}$, $S_{B,x}$, $S_{B,y}$, $S_{B,z}$, $S^{\prime}_{B,x}$, $S^{\prime}_{B,y}$, and $S^{\prime}_{B,z}$ are independent in each IR. }
\begin{ruledtabular}
\begin{tabular}{ccccccc}
IR & Cu1 & Cu2 & Cr1 & Cr2 & Cr3 & Cr4 \\
  & (0, 1/4, 3/8) & (0, 3/4, 5/8) & (0, 0, 0) & (0, 1/2, 0) & (1/4, 1/4, 3/4) & (1/4, 3/4, 1/4) \\
\hline
$\Gamma_{1}$ $A_{1g}$ & -- & -- & ($S_{B,x}$, 0, 0) & (-$S_{B,x}$, 0, 0) & (0, -$S_{B,x}$, 0) & (0, $S_{B,x}$, 0) \\
$\Gamma_{3}$ $A_{2g}$ & (0, 0, $S_{A,z}$) & (0, 0, $S_{A,z}$) & (0, $S_{B,x}$, $S_{B,z}$) & (0, -$S_{B,x}$, $S_{B,z}$) & ($S_{B,x}$, 0, $S_{B,z}$) & (-$S_{B,x}$, 0, $S_{B,z}$) \\
$\Gamma_{5}$ $B_{1g}$ & -- & -- & ($S_{B,x}$, 0, 0) & (-$S_{B,x}$, 0, 0) & (0, $S_{B,x}$, 0) & (0, -$S_{B,x}$, 0) \\
$\Gamma_{7}$ $B_{2g}$ & -- & -- & (0, $S_{B,x}$, $S_{B,z}$) & (0, -$S_{B,x}$, $S_{B,z}$) & (-$S_{B,x}$, 0, -$S_{B,z}$) & ($S_{B,x}$, 0, -$S_{B,z}$) \\
$\Gamma_{9}$ $E_{g}$ & ($S_{A,x}$, $S_{A,y}$, 0) & ($S_{A,x}$, $S_{A,y}$, 0) & ($S_{B,x}$, $S_{B,y}$, $S_{B,z}$) & ($S_{B,x}$, $S_{B,y}$, -$S_{B,z}$) & (-$S^{\prime}_{B,x}$, -$S^{\prime}_{B,y}$, -$S^{\prime}_{B,z}$) & (-$S^{\prime}_{B,x}$, -$S^{\prime}_{B,y}$, $S^{\prime}_{B,z}$) \\
\end{tabular}
\end{ruledtabular}
\end{table*}
\begin{table*}[htbp]
\caption{\label{tab:BVs_011K} Basis vectors of magnetic structures represented by $\Gamma_{\rm Cu-Cr}$ used for the Mag-I phase. The space group is described by $Fddd$ (No.~70, origin 2) and the magnetic propagation vector $\vec{k}=(0, 0, 0)$. The parameters of $S_{A,x}$, $S_{A,y}$, $S_{A,z}$, $S_{B,x}$, $S_{B,y}$, and $S_{B,z}$ are independent in each IR. }
\begin{ruledtabular}
\begin{tabular}{ccccccc}
IR & Cu1 & Cu2 & Cr1 & Cr2 & Cr3 & Cr4 \\
  & (1/8, 1/8, 1/8) & (7/8, 7/8, 7/8) & (1/2, 1/2, 1/2) & (1/2, 1/4, 1/4) & (1/4, 1/2, 1/4) & (1/4, 1/4, 1/2) \\
\hline
$\Gamma_{3}$ $B_{3g}$ & ($S_{A,x}$, 0, 0) & ($S_{A,x}$, 0, 0) & ($S_{B,x}$, $S_{B,y}$, $S_{B,z}$) & ($S_{B,x}$, -$S_{B,y}$, -$S_{B,z}$) & ($S_{B,x}$, -$S_{B,y}$, $S_{B,z}$) & ($S_{B,x}$, $S_{B,y}$, -$S_{B,z}$) \\
$\Gamma_{5}$ $B_{2g}$ & (0, $S_{A,y}$, 0) & (0, $S_{A,y}$, 0) & ($S_{B,x}$, $S_{B,y}$, $S_{B,z}$) & (-$S_{B,x}$, $S_{B,y}$, $S_{B,z}$) & (-$S_{B,x}$, $S_{B,y}$, -$S_{B,z}$) & ($S_{B,x}$, $S_{B,y}$, -$S_{B,z}$) \\
$\Gamma_{7}$ $B_{1g}$ & (0, 0, $S_{A,z}$) & (0, 0, $S_{A,z}$) & ($S_{B,x}$, $S_{B,y}$, $S_{B,z}$) & (-$S_{B,x}$, $S_{B,y}$, $S_{B,z}$) & ($S_{B,x}$, -$S_{B,y}$, $S_{B,z}$) & (-$S_{B,x}$, -$S_{B,y}$, $S_{B,z}$) \\
\end{tabular}
\end{ruledtabular}
\end{table*}

\clearpage
\begin{figure}[htbp]
\begin{center}
\includegraphics[width=0.4\linewidth, keepaspectratio]{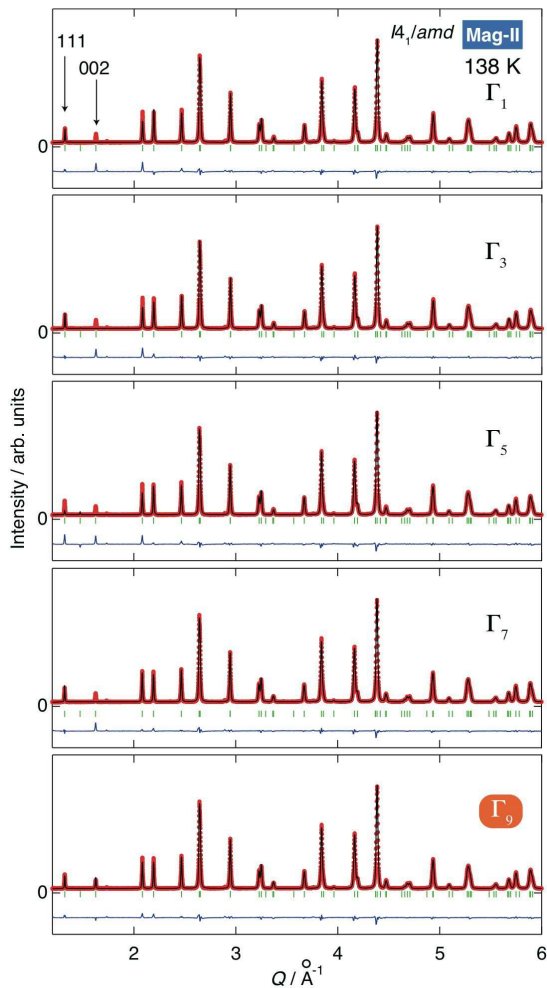}
\end{center}
\caption{\label{fig:SI_neutron_138K} 
Rietveld-fitting results of the $\Gamma_{\rm Cr}$ IR magnetic structures obtained for the 138-K neutron diffraction data. }
\end{figure}
%

%
\subsection{II. Details of structural parameters}
The structural parameters refined by Rietveld analysis are summarized in Table~\ref{tab:rietveld}. Here, by using these parameters, we discuss the effect of tetragonal Jahn-Teller deformation. 

First, the Cu-O distance in the CuO$_4$ tetrahedron, defined by $d$(Cu-O), is identical for the four Cu-O bonds. Thus, in order to discuss the CuO$_4$ deformation, we define the parameters of $d$(Cu-O$_{{\rm av},a}$), $d$(Cu-O$_{{\rm av},b}$), and $d$(Cu-O$_{{\rm av},c}$), where O$^{\rm av}_{a}$, O$^{\rm av}_{b}$, and O$^{\rm av}_{c}$ denote the midpoints between two oxygen atoms away from Cu along the $a$-, $b$-, and $c$-axes in CuO$_4$, respectively, as shown in Fig.~3(c) in the main text. The $d$(Cu-O$^{\rm av}_{c}$) is certainly much shorter than $d$(Cu-O$^{\rm av}_{a}$) and $d$(Cu-O$^{\rm av}_{b}$), as the source of Jahn-Teller lattice deformation. 
In addition, similarly, the out-of-plane first-neighbor Cr-Cr distances, defined by $d$(Cr-Cr)$_{bc}$ and $d$(Cr-Cr)$_{ca}$, are also much shorter than the in-plane distance, defined by $d$(Cr-Cr)$_{ab}$. This supports the notion of $|J_{1c}| > |J_{1ab}|$ as discussed in the main text. 

By contrast, the Cr-O distances along the $a$-, $b$-, and $c$-axes in CrO$_6$, defined by $d$(Cr-O$_a$), $d$(Cr-O$_b$), and $d$(Cr-O$_c$), respectively, are less deformed, of which the change ratio is only $\Delta R_{\rm out}$(Cr-O) $\sim 10^{-3}$ in this order. This fact indicates that, interestingly, the tetragonal Jahn-Teller splitting energy inside the $d\epsilon$ orbital in Cr is much smaller than that in Cu: $\Delta_{\rm Cr,JT} \ll \Delta_{\rm Cu,JT}$. Therefore, a further enhancement of the second-order SOC mechanism is expected for Cr. On the other hand, unlike the Cu hole, the Cr hole is not explicit but is generated to a slight extent by the direct AF $J_{1c}$ exchange process, which suppresses the mechanism. Thus, we consider that, as a consequence of competition between these two factors, the moderate second-order SOC effects of unquenched $L$ and CrO$_6$ deformation would have been observed. 

%
\begin{table*}[htbp]
\caption{\label{tab:rietveld} Structural parameters refined by Rietveld analysis. The atomic positions are described by Cu $4a$ (0, 1/4, 3/8), Cr $8d$ (0, 0, 0), O $16h$ $(0, y, z)$ in $I4_{1}/amd$, and Cu $8a$ (1/8, 1/8, 1/8), Cr $16d$ (1/2, 1/2, 1/2), O $32h$ $(x, y, z)$ in $Fddd$. The superscript $I$ denotes the use of $I$-lattice notation for convenience. 
The parameters of distance $d$ are defined in the text. The parameter $M$ denotes the ordered magnetic moment. }
\begin{ruledtabular}
\begin{tabular}{cccc}
 & 199 K ($I4_{1}/amd$) & 138 K ($I4_{1}/amd$) & 11 K ($Fddd$) \\
\hline
$a$ / {\AA} & 6.03331(1)$^I$ & 6.03630(2)$^I$ & 8.53970(6) \\
$b$ / {\AA} & 6.03331(1)$^I$ & 6.03630(2)$^I$ & 8.54435(6) \\
$c$ / {\AA} & 7.75808(3)$^I$ & 7.73557(4)$^I$ & 7.71225(3) \\
$x$ (O) & 0$^I$ & 0$^I$ & 0.26828(22) \\ 
$y$ (O) & 0.53500(10)$^I$ & 0.53532(12)$^I$ & 0.26729(21) \\
$z$ (O) & 0.25383(8)$^I$ & 0.25457(10)$^I$ & 0.24471(7) \\
\hline
$d$(Cu-O) / {\AA} & 1.9597(7) & 1.9581(8) & 1.9566(15) \\
$d$(Cu-O$^{\rm av}_{a}$) / {\AA} & 1.2159(3) & 1.2173(3) & 1.2237(11) \\
$d$(Cu-O$^{\rm av}_{b}$) / {\AA} & 1.2159(3) & 1.2173(3) & 1.2159(11) \\
$d$(Cu-O$^{\rm av}_{c}$) / {\AA} & 0.9400(5) & 0.9316(6) & 0.9232(4) \\
$d$(Cr-Cr)$_{ab}$ / {\AA} & 3.016657(7) & 3.018151(9) & 3.020062(15) \\
$d$(Cr-Cr)$_{bc}$ / {\AA} & 2.883028(6) & 2.880028(7) & 2.877550(12) \\
$d$(Cr-Cr)$_{ca}$ / {\AA} & 2.883028(6) & 2.880028(7) & 2.876687(12) \\
$d$(Cr-O$_a$) / {\AA} & 1.9896(4) & 1.9894(5) & 1.9846(17) \\
$d$(Cr-O$_b$) / {\AA} & 1.9896(4) & 1.9894(5) & 1.9948(17) \\
$d$(Cr-O$_c$) / {\AA} & 1.9805(7) & 1.9808(9) & 1.9806(6) \\
\hline
$M_a$ (Cu) / $\mu_{\rm B}$ & - & 0 & -0.87(3) \\
$M_a$ (Cr) / $\mu_{\rm B}$ & - & 0 & 1.22(2) \\
$M_b$ (Cr) / $\mu_{\rm B}$ & - & 1.56(2) & 2.68(1) \\
\end{tabular}
\end{ruledtabular}
\end{table*}
%

%
%

%
\subsection{III. Transition temperature of tetragonal-to-orthorhombic lattice deformation}
We complementarily verified this by neutron diffraction, which was reported as $T_{\rm I}$ rather than $T_{\rm II}$ by using synchrotron X-ray diffraction~\cite{Suchomel_2012}. Figure~\ref{fig:SI_Tdep040} shows the temperature evolution of the 040 and 400 reflections. The line profile is described by a single Lorentzian in the Para phase and by a double Lorentzian in the Mag-I phase. In the Mag-II phase, the single Lorentzian sufficiently fits the experimental data without requiring the double Lorentzian. Thus, neutron diffraction shows orthorhombic lattice deformation to occur at $T_{\rm I}$, in harmony with X-ray diffraction. 

\begin{figure}[htbp]
\begin{center}
\includegraphics[width=0.38\linewidth, keepaspectratio]{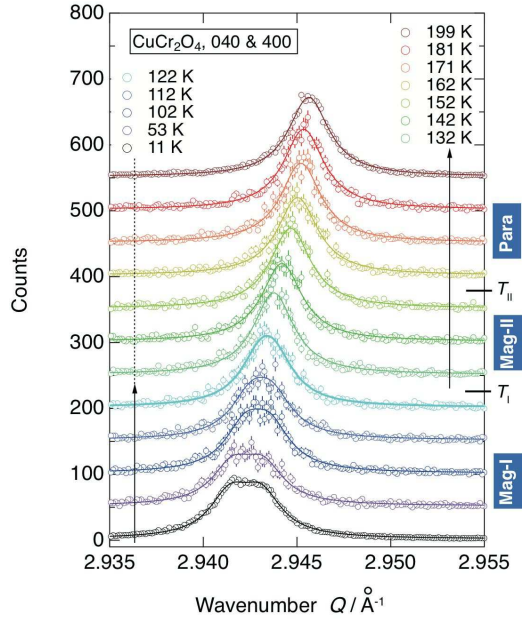}
\end{center}
\caption{\label{fig:SI_Tdep040} 
Temperature evolution of the 040 and 400 reflections recorded in high-resolution backward bank. }
\end{figure}
%

%
\section{IV. Magnetization curves}

Figure~\ref{fig:SI_MH} shows the measured magnetization curves. Overall, spontaneous magnetization accompanied by hysteresis is observed in the low-temperature range, indicating growing ferrimagnetism, as shown in (a). As the temperature decreases from 200 K (Para phase), the remanent magnetization is virtually zero but becomes slightly finite below $T_{\rm II}\simeq155$ K, and rapidly increases below $T_{\rm I}\simeq125$ K. These behaviors are clearly seen in the magnifications in (b) and (c), respectively. 
Thus, the two magnetic transitions at $T_{\rm II}$ and $T_{\rm I}$ are also confirmed in the magnetization curves. 

\begin{figure*}[htbp]
\begin{center}
\includegraphics[width=0.8\linewidth, keepaspectratio]{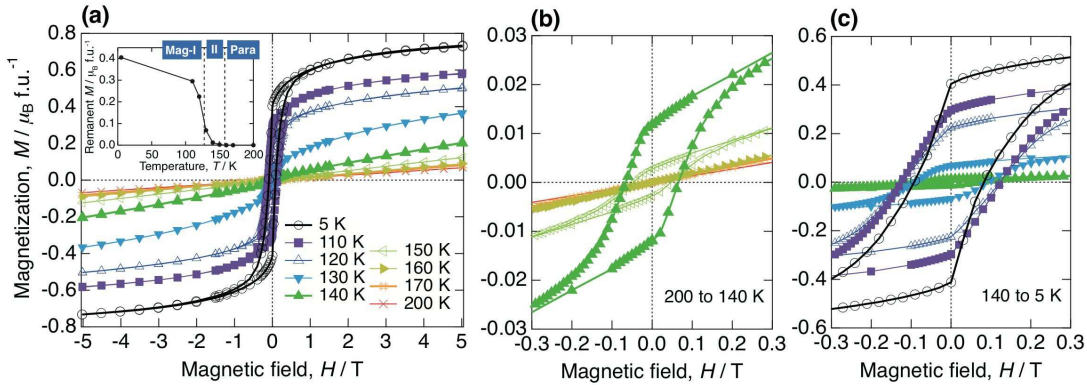}
\end{center}
\caption{\label{fig:SI_MH} 
(a) Magnetization process measured at several temperatures. The unit f.u.~denotes the formula unit CuCr$_2$O$_4$. The inset shows the temperature dependence of remanent magnetization values. (b) Low-field and low-magnetization range magnified from 200 to 140 K. (c) Low-field range magnified from 140 to 5 K. For clarity, the 140-K data is shown common to both (b) and (c). In all these figures, the lines are intended to guide the eye. }
\end{figure*}

We compare the magnetization data to the magnetic structures determined by neutron diffraction. 
First, for the Mag-II phase, we note that the 140-K spontaneous magnetization of approximately 0.02 $\mu_{\rm B}$/f.u. is too weak to detect by neutron diffraction, which is consistent with the basic AF magnetic structure in the Mag-II phase determined by the 138-K neutron diffraction data. Furthermore, the existence of spontaneous magnetization is symmetrically allowed, as listed in the $\Gamma_{9}$ array in Tab.~\ref{tab:BVs_138K}.  
Second, for the Mag-I phase, by extrapolating the 5-K magnetization curve from 4 -- 5 Tesla to zero field, the powder-averaged spontaneous magnetization is roughly estimated as 0.7 $\mu_{\rm B}$/f.u. On the other hand, neutron diffraction provides the values of $\{ 1\cdot M_{a}({\rm Cu}) + 2\cdot M_{a}({\rm Cr}) \} /2 = \{ 1\cdot(-0.87) + 2\cdot1.22 \} /2 = 0.8$ $\mu_{\rm B}$/f.u.~at 11 K, where the division by 2 denotes a typical factor of powder averaging~\cite{Book_Chikazumi_1997}. 
Thus, the results of magnetization and neutron diffraction are excellently consistent with each other within approximately 0.1 $\mu_{\rm B}$/f.u.~deviation in both the Mag-II and Mag-I phases. However, the further precise modification of the magnetic structure by this very small component in the Mag-II phase is beyond the scope of this study and could be investigated in future, for example, by the complementary techniques of nuclear magnetic resonance and soft-X-ray scattering.

\section{V. Thermal conductivity}

The mean free path ($\lambda$) of phonons obtained from the thermal conductivity ($\kappa$) is useful to examine the existence of spin-lattice coupling in a system~\cite{Zhou_2013}. 
The equation $\kappa=C_{p}v\lambda$ is used to estimate the information of $\lambda(T)$. Here, $v$ denotes the velocity of sound, which is normally independent from temperature, i.e., approximately 10{\%} at most even in frustrated MgCr$_{2}$O$_{4}$ accompanied by strong spin-lattice coupling~\cite{Watanabe_2012}. Hence, normalized $\lambda(T)/\lambda(T_{0}) = \{ \kappa(T)/\kappa(T_{0}) \} / \{ C_{p}(T)/C_{p}(T_{0}) \} $ is obtained from the measured $\kappa(T)$ and $C_{p}(T)$. 

Thus, we measured the temperature dependence of the thermal conductivity using the Physical Properties Measurement System (PPMS) at the Department of Applied Physics, Tohoku University. Figure~\ref{fig:SI_thermalC} shows the measured $\kappa(T)/\kappa(T_{0})$ and the resultant $\lambda(T)/\lambda(T_{0})$. The ratio $\lambda(T)/\lambda(T_{0})$ rapidly increases by about two orders with decreasing temperature from the Para and Mag-II phase to the Mag-I phase, which is consistent with the fact that the temperature dependence of $v$ is negligible. 

\begin{figure}[htbp]
\begin{center}
\includegraphics[width=0.4\linewidth, keepaspectratio]{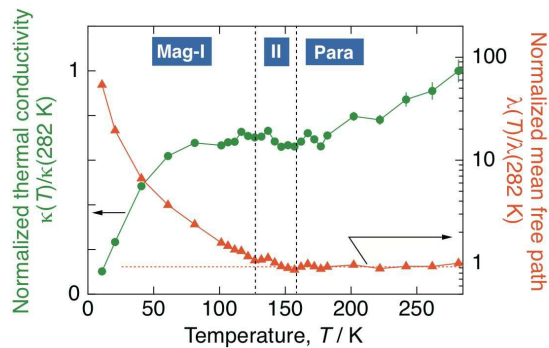}
\end{center}
\caption{\label{fig:SI_thermalC} 
Temperature dependence of thermal conductivity (left, green) and mean free path (right, orange). Both values are normalized to be 1 at 282 K. }
\end{figure}

This significant change in $\lambda(T)/\lambda(T_{0})$ warrants discussion. In amorphous materials, disorder or short-range order causes thermal insulation~\cite{Kittel_1949}. A similar effect is also observed in highly spin-frustrated MgCr$_{2}$O$_{4}$ and ZnCr$_{2}$O$_{4}$, in which the dynamical short-range order of molecular spin excitations hinders phonon travel through spin-lattice coupling~\cite{Watanabe_2012, Zhou_2013}. In analogy with these, the following understanding is suggested for CuCr$_2$O$_4$. In the Para and Mag-II phases, the substantial component of thermally fluctuating magnetic disorder and/or short-range order hinder the phonon travel, as the Cr ordered moment is 1.56$\mu_{\rm B}$, which is much less than the full value of 3$\mu_{\rm B}$. As the temperature decreases to the Mag-I phase, the amorphous-like component decreases, the static magnetic long-range order grows, and hence the thermal-insulation effect disappears. 
Thus, this thermal-conductivity result experimentally supports the existence of the spin-lattice coupling in CuCr$_2$O$_4$. 

Notably, the Mag-II phase exhibits substantial spin-lattice fluctuation as the microscopic origin for the macroscopic physical property of thermal insulation. In accordance with the observed long-range order of the Cr collinear $uudd$ type, the fluctuation probably also forms the $uudd$ unit locally, which was theoretically predicted to compose the spin nematic state~\cite{Shannon_2010}. We expect further studies of the Mag-II phase to be promising towards its realization. \\

\end{widetext}

\bibliography{CuCr2O4_3_arXiv}

\end{document}